\documentclass[lettersize,journal]{IEEEtran}

\usepackage{graphicx}	
\usepackage{amsmath}	
\usepackage{amssymb}	
\usepackage{booktabs}
\usepackage{times}
\usepackage{microtype}
\usepackage{epsfig}
\usepackage[caption=false,font=normalsize,labelfont=sf,textfont=sf]{subfig}
\usepackage{float}
\usepackage{placeins}
\usepackage{color}
\usepackage{colortbl}
\usepackage{stfloats}
\usepackage{enumitem}
\usepackage{tabularx}
\usepackage{multirow}
\usepackage{xspace}
\usepackage{url}
\usepackage{xcolor}
\usepackage{arydshln}
\usepackage{cite}
\usepackage{hyperref}
\hypersetup{hypertex=true,
colorlinks=true,
linkcolor=blue,
anchorcolor=blue,
citecolor=blue}

\def\BibTeX{{\rm B\kern-.05em{\sc i\kern-.025em b}\kern-.08em
    T\kern-.1667em\lower.7ex\hbox{E}\kern-.125emX}}

\begin{document}
\title{LKFMixer: Exploring Large Kernel Feature For Efficient Image Super-Resolution}
\author{Yinggan Tang, Quanwei Hu
\thanks{Yinggan Tang and Quanwei Hu are with the School of Electrical Engineering, Yanshan University, Qinhuangdao, Hebei 066004, China. (Email: ygtang@ysu.edu.cn, quanwei1277@163.com)
\par
Corresponding author: Yinggan Tang.}}

\maketitle

\begin{abstract}
The success of self-attention (SA) in Transformer demonstrates the importance of non-local information to image super-resolution (SR), but the huge computing power required makes it difficult to implement lightweight models. To solve this problem, we propose a pure convolutional neural network (CNN) model, LKFMixer, which utilizes large convolutional kernel to simulate the ability of self-attention to capture non-local features. Specifically, we increase the kernel size to 31 to obtain the larger receptive field as possible, and reduce the parameters and computations by coordinate decomposition. Meanwhile, a spatial feature modulation block (SFMB) is designed to enhance the focus of feature information on both spatial and channel dimension. In addition, by introducing feature selection block (FSB), the model can adaptively adjust the weights between local features and non-local features. Extensive experiments show that the proposed LKFMixer family outperform other state-of-the-art (SOTA) methods in terms of SR performance and reconstruction quality. In particular, compared with SwinIR-light on Manga109 dataset, LKFMixer-L achieves 0.6dB PSNR improvement at $\times$4 scale, while the inference speed is $\times$5 times faster. The code is available at \href{https://github.com/Supereeeee/LKFMixer}{https://github.com/Supereeeee/LKFMixer}.
\end{abstract}


\section{Introduction}
\label{sec:intro}
\IEEEPARstart{T}{he} process of recovering its high-resolution (HR) counterpart from a low-resolution (LR) image is called image super-resolution (SR). Recently, the excellent performance of SR models based on convolutional neural networks (CNNs) benefits from the increased network complexity \cite{He2016, Lim2017}. However, these models face challenges when deployed in resource-constrained scenarios, which stimulates people to develop lightweight SR models with fewer parameters, higher efficiency, and faster inference speed \cite{Ren2024}. With reduced model complexity, the performance of lightweight SR models highly depends on efficient feature extraction and large receptive field \cite{Lee2024}. Nevertheless, the localization of convolution operations in CNN-based SR models lead to limited receptive field though it has powerful local feature extraction ability. In contrast, the self-attention (SA) mechanism in Transformer-based SR models \cite{Dosovitskiy2021} possesses powerful ability to capture non-local features and long-range dependencies, but this comes at the cost of high computational resource demands and slow inference speed. Is there a way that makes CNNs equip with Transformer's ability of capturing non-local features with less computation consumption and inference time? One potential solution may lie in the adoption of larger convolution kernels.
\begin{figure}[!ht]
\centering
\includegraphics[width=1\columnwidth]{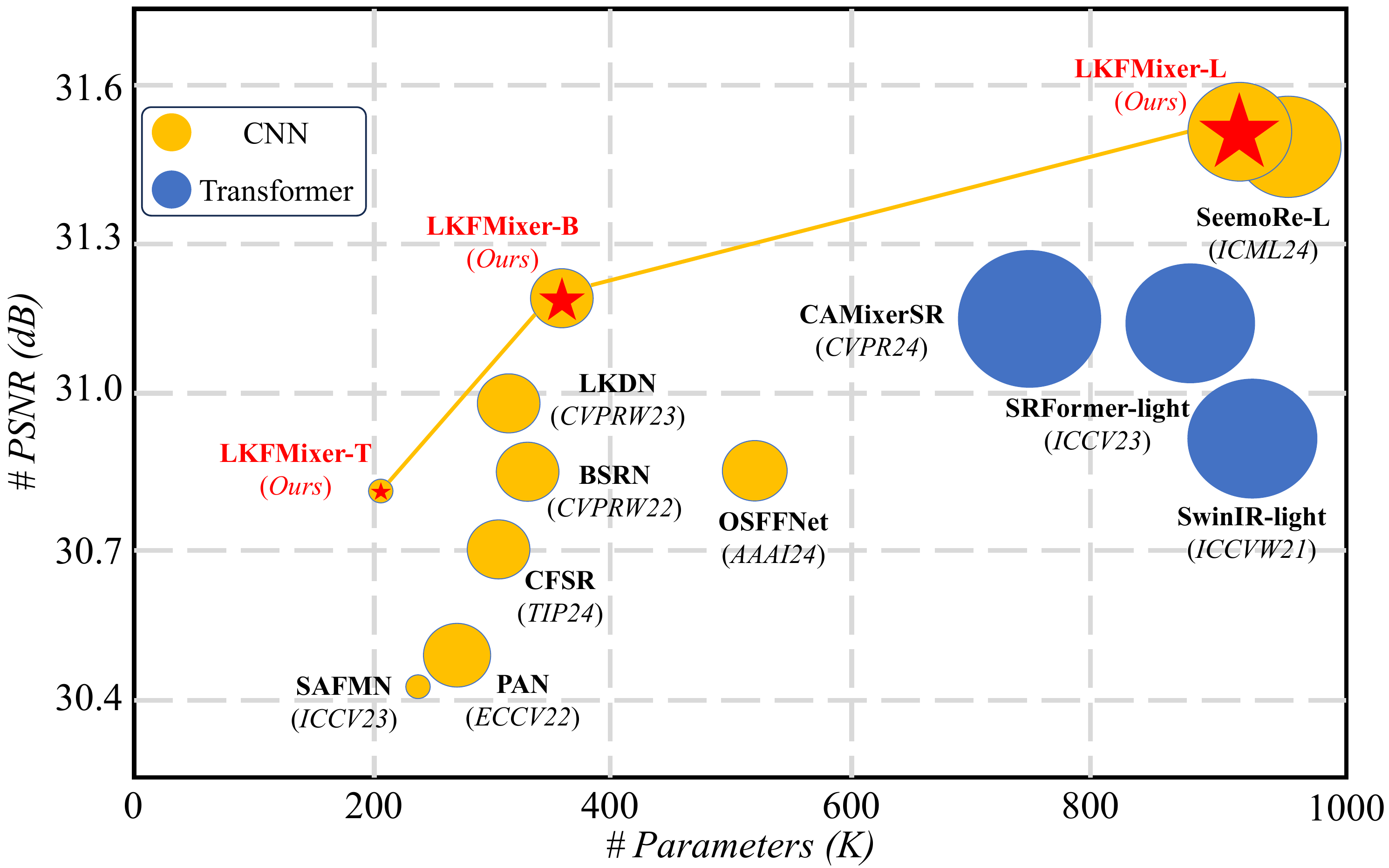}
\caption{SR performance and model complexity comparison between our proposed LKFMixer family and other SOTA lightweight models on Manga109 for $\times$4 SR task. The circle sizes indicate the number of FLOPs.}
\label{fig:Model complexity and performance}
\end{figure}
Previous studies \cite{Liu2022, Ding2022} have demonstrated that increasing the size of the convolution kernel can effectively expand the receptive field, thereby enhancing the extraction of non-local features. However, this also leads to a significant increase in model complexity as the model complexity is proportional to the square of the convolution kernel size. To mitigate this issue and reduce the model complexity of CNNs with large convolution kernels, several approaches have been adopted, such as dilated convolution \cite{Song2024} and large kernel decomposition \cite{Guo2023}. However, compared with the true large kernel convolutions, these methods often fail to effectively utilize all pixels in the same receptive field and result in feature information loss to a certain extent, which will negatively impact the SR performance and the quality of image reconstruction.
\par
To address the above issues, we propose a novel method called LKFMixer, which applies large convolution kernel in CNN-based lightweight SR models for the purpose of enlarging receptive field while reducing model complexity. To be specific, the size of depth-wise convolution (DWConv) kernel is increased to 31, which is decomposed into DWConv1$\times$31 and DWConv31$\times$1 in series by coordinate decomposition. To further reduce information redundancy across channels and model complexity, we apply partial convolution (PConv) \cite{Chen2023}, utilizing the decomposed large convolution kernel to extract features from the first quarter of the channels. Based on above design, we propose a partial large kernel block (PLKB), which allows for an increased kernel size with little impact on model complexity and inference time, making it well-suited for lightweight SR models. Additionally, we introduce a feature fusion block (FFB) that aggregates the outputs of DWConv3$\times$3 and PLKB to extract both local and non-local features. By utilizing FFB, we construct a feature distillation block (FDB) as the core unit for feature extraction and fusion. Meanwhile, we introduce a feature selection block (FSB) to adaptively balance the contribution of local features and non-local features. In addition, in order to pay more attention to spatial and channel information, a spatial feature modulation block (SFMB) is also proposed to improve the SR performance.
\par
As shown in Fig. \ref{fig:Model complexity and performance}, the proposed LKFMixer family strikes a better trade-off between model complexity and SR performance compared with the recent state-of-the-art (SOTA) lightweight SR model. In summary, our contributions are as follows
\begin{itemize}
\item We propose PLKB, which effectively captures non-local features and alleviates the increase in model complexity brought by large convolution kernel.
\item By introducing SFMB and FSB, our model adaptively adjusts the contribution of local and non-local features, while simultaneously enhancing spatial and channel information.
\item Extensive experiments demonstrate that our proposed LKFMixer family outperforms current SOTA lightweight SR models, achieving superior SR performance while maintaining faster inference speed.
\end{itemize}
\section{Related work}
\label{Related work}
\subsection{CNN-based Efficient SR}
\label{subsec:CNN based Efficient SR}
Thanks to the powerful feature extraction capability of CNNs, numerous lightweight SR models based on CNNs have rapidly emerged since the introduction of SRCNN \cite{Dong2014}. To reduce model parameters, various variants of standard convolution, such as depth-wise convolution (DWConv) \cite{Chollet2017} and blueprint separable convolution (BSConv) \cite{Haase2020}, have been introduced into lightweight SR models. Meanwhile, feature distillation strategy has been employed in IMDN \cite{Hui2019}, RFDN \cite{Liu2020}, and BSRN \cite{Li2022} to refine features. Additionally, re-parameterization technique has also been adopted in RepRFN \cite{Deng2023} to accelerate inference. In terms of attention mechanism, enhanced spatial attention (ESA) and contrast-aware channel attention (CCA) have been proposed in RFANet \cite{Liu2020a} and IMDN \cite{Hui2019} to improve SR performance with fewer model parameters, respectively. In this paper, we explore the potential of applying large convolution kernel to lightweight SR model.
\subsection{Transformer-based Efficient SR}
\label{subsec:Transformer based Efficient SR}
Transformer \cite{Dosovitskiy2021} exhibits excellent property in modeling long-distance dependencies but suffers from the substantial computational burden imposed by the self-attention (SA) mechanism. To reduce the model complexity, SwinIR \cite{Liang2021} introduces sliding local windows to calculate attention weights. ELAN \cite{Zhang2022} employs group self-attention to improve efficiency. SRFormer \cite{Zhou2023} leverages permuted self-attention to enjoy the benefit of large window self-attention. CAMixerSR \cite{Wang2024a} combines convolution with deformable window self-attention to achieve superior SR performance. While these SR models effectively reduce the complexity of self-attention, a significant gap remains in actual inference speed when compared with CNN-based SR models.
\subsection{Large kernel design}
\label{subsec:Large kernel design}
Currently, researchers prefer to stack multiple 3$\times$3 convolutions to reduce model parameters. However, it has shown that larger convolution kernels can capture more feature information and significantly expand the receptive field. For example, ConvNeXt \cite{Liu2022} employs 7$\times$7 convolution and achieves the same even better performance compared with Swin-Transformer. RepLKNet \cite{Ding2022} suggests guidelines for large kernel design and increases the convolution kernel size to 31$\times$31. PeLK \cite{Chen2024} introduces a human-like peripheral convolution, further increasing the kernel size to an impressive 101$\times$101. These models all highlight the effectiveness of large convolution kernels. However, the substantial computations required by large convolution kernel remain a significant challenge, which will be further explored in this paper.
\section{Proposed method}
\label{Proposed method}
\begin{figure*}[!ht]
\centering
\includegraphics[width=1\linewidth]{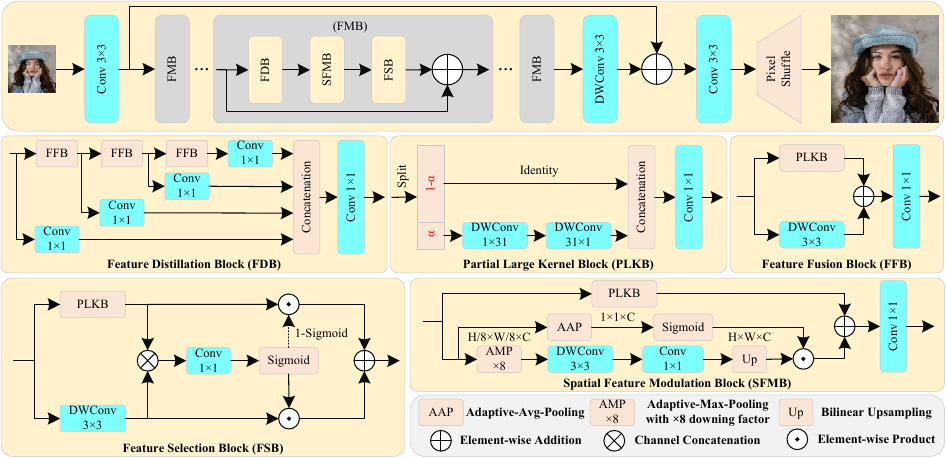}
\caption{The overall architecture of LKFMixer, and the detail structure of FDB, SFMB and FSB.}
\label{fig:The overall architecture of LKFMixer}
\end{figure*}

In this section, we provide a detailed overview of the various components of the proposed model. Fig. \ref{fig:The overall architecture of LKFMixer} illustrates the overall architecture of LKFMixer, which is primarily composed of three key components. The first  component is a 3$\times$3 convolution layer for shallow feature extraction. The second component consists of multiple stacked feature modulation blocks (FMBs) for deep feature extraction. Each FMB is composed of a sequentially arrangement of a feature distillation block (FDB), a spatial feature modulation block (SFMB), and a feature selection block (FSB). The third component is the up-sampler module, which includes a 3$\times$3 convolutional layer followed by a sub-pixel layer \cite{Shi2016}.
\subsection{Feature Distillation Block}
\label{subsec:Feature Distillation Block}
\subsubsection{Partial Large Kernel Block}
\label{subsubsec:Partial Large Kernel Block}
FasterNet \cite{Chen2023} proposed partial convolution (PConv) that performs convolution operations only on partial channels. Inspired by PConv, we extract non-local features by adopting DWConv31$\times$31 from the input $\rm{I}_{in}$ on partial channels while the remaining channels keep unchanged. Furthermore, we decompose DWConv31$\times$31 into two stripe convolutions in series, DWConv1$\times$31 and DWConv31$\times$1, via coordinate decomposition to further reduce model complexity. This process can be expressed as
\begin{equation}
\rm{I}_{out}^{\alpha C} = DWConv_{31\times1}(DWConv_{1\times31}(I_{in}^{\alpha C}))
\end{equation}
where $\alpha$ denotes the split factor of the channels, $C$ denotes the number of channels of $\rm{I}_{in}$, $\rm{I}_{in}^{\alpha C}$ represents the first $\alpha C$ channels of $\rm{I}_{in}$, and $\rm{I}_{out}^{\alpha C}$ indicates the output of decomposed large kernel on $\alpha C$ channels. Then, 1$\times$1 convolution is utilized for information fusion after concatenation along the channel dimension. The final output $\rm{I}_{out}$ of PLKB can be calculated as
\begin{equation}
\rm{I}_{out} = Conv_{1\times1}(I_{out}^{\alpha C}\otimes I_{in}^{(1-\alpha)C})
\end{equation}
where $\otimes$ indicates the channel concatenation operation, and $\rm{I}_{in}^{(1-\alpha)C}$ denotes the features of remaining unchanged channels.\\
\subsubsection{Feature Fusion Block}
\label{subsubsec:Feature Fusion Block}
We extract local and non-local features via DWConv3$\times$3 and PLKB, respectively. A 1$\times$1 convolution layer is adopted for information fusion. For the input $\rm{I}_{in}$ and output $\rm{I}_{out}$ of FFB, the calculation process can be described as
\begin{equation}
\rm{I}_{out} = Conv_{1\times1}(DWConv_{3\times3}(I_{in})+F_{PLKB}(I_{in}))
\end{equation}
By embedding multiple FFBs into various levels of the distillation structure, rich local and non-local features are gradually refined in the trunk branch. To eliminate redundant information, 1$\times$1 convolutions are used in the distillation branch. The whole process of FDB can be expressed as
\begin{equation}
\begin{aligned}
\rm{I}_{r}^{1},I_{d}^{1}&=\rm{FFB}_{1}(I_{in}),Conv_{1\times1}^{1}(I_{in})\\
\rm{I}_{r}^{2},I_{d}^{2}&=\rm{FFB}_{2}(I_{r}^{1}),Conv_{1\times1}^{2}(I_{r}^{1})\\
\rm{I}_{r}^{3},I_{d}^{3}&=\rm{FFB}_{3}(I_{r}^{2}),Conv_{1\times1}^{3}(I_{r}^{2})\\
\end{aligned}
\end{equation}
where $\rm{I}_{in}$ is the input, $\rm{I}_{r}^{\it{i}}$ and $\rm{I}_{d}^{\it{i}}$ are the $i$th layer output of refine branch and distillation branch. $\rm{FFB}_{\it{i}}, \rm{Conv}_{1\times1}^{\it{i}}(\it{i}=1,2,3)$ denote the $i$th FFB and 1$\times$1 convolution, respectively. Finally, a 1$\times$1 convolution layer is utilized to merge the feature information after channel concatenation, expressed as
\begin{equation}
\rm{I}_{out}=Conv_{1\times1}(I_{d}^{1} \otimes I_{d}^{2} \otimes I_{d}^{3} \otimes I_{r}^{3})\\
\end{equation}
where $\rm{I}_{out}$ is the output of FDB, and $\otimes$ indicates the channel concatenation operation.
\subsection{Spatial Feature Modulation Block}
\label{subsec:Spatial Feature Modulation Block}
Both channel and spatial information are critical for SR task. To this end, we design a spatial feature modulation block (SFMB). It includes a spatial branch that captures low-frequency spatial features through downsampling, along with a channel attention branch that focuses on channel information. The computational process is given by
\begin{equation}
\rm{I}_{s}=\rm{F}_{up}(Conv_{1\times1}(F_{AMP}^{\times8}(I_{in}))) \odot F_{sig}(F_{AAP}(I_{in}))\\
\end{equation}
where $\rm{F}_{AAP}$ and $\rm{F}_{AMP}^{\times8}$ are the operation of Adaptive-Avg-Pooling and Adaptive-Max-Pooling with $\times$8 downing factor along the spatial dimension, respectively. $\rm{F}_{sig}$ denotes the sigmoid function, $\rm{F}_{up}$ indicates Bilinear upsampling operation, and $\odot$ denotes element-wise product. By utilizing a 1$\times$1 convolution to fuse the information from the spatial branch and the PLKB module, both non-local features and low-frequency spatial feature are extracted simultaneously. The process is expressed as
\begin{equation}
\rm{I}_{out}=Conv_{1\times1}(F_{PLKB}(I_{in})+I_{s})\\
\end{equation}
where $\rm{I}_{in}$ and $\rm{I}_{out}$ are the input and output of SFMB, $\rm{F}_{PLKB}$ is the PLKB module.
\subsection{Feature Selection Block}
\label{subsec:Feature Selection Block}
To adaptively adjust the weights of local and non-local features, we develop a feature selection block (FSB). First, the features from the PLKB and the DWConv3$\times$3 are concatenated along the channel dimension and aggregated by a 1$\times$1 convolution. The fused features go through a sigmoid function to obtain the attention weights $\beta$. The features of DWConv3$\times$3 branch are weighted by $\beta$ while the features of PLKB branch are weighted by $1-\beta$. The two branches influence each other, adjusting the weight between non-local and local features. Finally, the information from the two branches is aggregated via summation operation. The entire process is as follows
\begin{equation}
\begin{aligned}
\rm{I}_{PLKB}, &\rm{I}_{DWConv3}=\rm{F}_{PLKB}(\rm{I}_{in}),DWConv_{3\times3}(\rm{I}_{in})\\
\beta &=\rm{F}_{sig}(\rm{F}_{Conv1\times1}(\rm{I}_{PLKB} \otimes \rm{I}_{DWConv3}))\\
\rm{I}_{out}&=\rm{I}_{DWConv3} \odot \beta + I_{PLKB} \odot (1-\beta)\\
\end{aligned}
\end{equation}
where $\rm{I}_{in}$ and $\rm{I}_{out}$ are the input and output of SFB, $\otimes$ and $\odot$ denote channel concatenation and element-wise product, respectively.
\par
\begin{table*}[!ht]
\caption{Quantitative comparison between the proposed method and other lightweight SR methods on benchmark datasets. The best and second best results are highlighted in \textcolor{red}{red} and \textcolor{blue}{blue} respectively. \#Flops is measured by recovering a 1280$\times$720 HR image.}
\label{tab:Quantitative comparison results with other lightweight SR methods}
\centering
\resizebox{\linewidth}{!}{
\begin{tabular}{cccccccccc}
\toprule
\multirow{2}{*}{Scale} & \multirow{2}{*}{Method} & \multirow{2}{*}{Years} & \#Params & \#Flops & Set5 & Set14 & BSD100 & Urban100 & Manga109\\
& & & [K] & [G] & PSNR / SSIM & PSNR / SSIM & PSNR / SSIM & PSNR / SSIM & PSNR / SSIM\\
\midrule
\multirow{20}{*}{×2}
& PAN \cite{Zhao2020} & ECCV2020 & 261 & 70.6 & 38.00 / 0.9605 & 33.59 / 0.9181 & 32.18 / 0.8997 & 32.01 / 0.9273 & 38.70 / 0.9773\\
& SAFMN \cite{Sun2023} & ICCV2023 & 228 & 52 & 38.00 / 0.9605 & 33.54 / 0.9177 & 32.16 / 0.8995 & 31.84 / 0.9256 & 38.71 / 0.9771\\
& MAN-tiny \cite{Wang2024b} & CVPRW2024 & 134 & 30 & 37.93 / 0.9604 & 33.48 / 0.9171 & 32.13 / 0.8993 & 31.75 / 0.9250 & 38.58 / 0.9770\\
& SMFANet \cite{Zheng2024} & ECCV2024 & 186 & 41 & \textcolor{blue}{38.08} / 0.9607 & 33.65 / 0.9186 & 32.22 / 0.9002 & 32.20 / 0.9282 & \textcolor{blue}{39.11} / \textcolor{blue}{0.9779}\\
& SeemoRe-T \cite{Zamfir2024} & ICML2024 & 220 & 46.1 & 38.06 / \textcolor{blue}{0.9608} & \textcolor{blue}{33.65} / \textcolor{blue}{0.9186} & \textcolor{blue}{32.23} / \textcolor{blue}{0.9004} & \textcolor{blue}{32.22} / \textcolor{blue}{0.9286} & 39.01 / 0.9777\\
\rowcolor{gray!20}
\cellcolor{white!} & \textbf{LKFMixer-T} & Ours & 194 & 41.1 & \textcolor{red}{\textbf{38.09}} / \textcolor{red}{\textbf{0.9609}} & \textcolor{red}{\textbf{33.73}} / \textcolor{red}{\textbf{0.9190}} & \textcolor{red}{\textbf{32.24}} / \textcolor{red}{\textbf{0.9007}} & \textcolor{red}{\textbf{32.30}} / \textcolor{red}{\textbf{0.9300}} & \textcolor{red}{\textbf{39.11}} / \textcolor{red}{\textbf{0.9781}}\\
\cdashline{2-10}
& ShuffleMixer \cite{Sun2022} & NIPS2022 & 394 & 91 & 38.01 / 0.9606 & 33.63 / 0.9180 & 32.17 / 0.8995 & 31.89 / 0.9257 & 38.83 / 0.9774\\
& BSRN \cite{Li2022} & CVPRW2022 & 332 & 72.2 & 38.10 / 0.9610 & 33.74 / 0.9193 & 32.24 / 0.9006 & 32.34 / 0.9303 & 39.14 / 0.9782\\
& VapSR \cite{Zhou2022} & ECCV2022 & 343 & 78.2 & 38.08 / \textcolor{blue}{0.9612} & 33.77 / 0.9195 & 32.27 / 0.9011 & 32.45 / 0.9316 & 39.09 / 0.9782\\
& LKDN \cite{Xie2023} & CVPRW2023 & 304 & 69.2 & \textcolor{blue}{38.12} / 0.9611 & \textcolor{blue}{33.90} / \textcolor{blue}{0.9202} & 32.27 / 0.9010 & 32.53 / 0.9322 & \textcolor{blue}{39.19} / \textcolor{blue}{0.9784}\\
& CFSR \cite{Wu2024} & TIP2024 & 291 & 62.6 & 38.07 / 0.9607 & 33.74 / 0.9192 & 32.24 / 0.9005 & 32.28 / 0.9300 & 39.00 / 0.9778\\
& OSFFNet \cite{Wang2024} & AAAI2024 & 516 & 83.2 & 38.11 / 0.9610 & 33.72 / 0.9190 & \textcolor{blue}{32.29} / \textcolor{blue}{0.9012} & \textcolor{blue}{32.67} / \textcolor{blue}{0.9331} & 39.09 / 0.9780\\
\rowcolor{gray!20}
\cellcolor{white!}& \textbf{LKFMixer-B} & Ours & 357 & 75.9 & \textcolor{red}{\textbf{38.21}} / \textcolor{red}{\textbf{0.9613}} & \textcolor{red}{\textbf{33.94}} / \textcolor{red}{\textbf{0.9209}} & \textcolor{red}{\textbf{32.33}} / \textcolor{red}{\textbf{0.9017}} & \textcolor{red}{\textbf{32.75}} / \textcolor{red}{\textbf{0.9337}} & \textcolor{red}{\textbf{39.29}} / \textcolor{red}{\textbf{0.9786}}\\
\cdashline{2-10}
& SwinIR-light \cite{Liang2021} & ICCVW2021 & 910 & 244 & 38.14 / 0.9611 & 33.86 / 0.9206 & 32.31 / 0.9012 & 32.76 / 0.9340 & 39.12 / 0.9783\\
& ELAN-light \cite{Zhang2022} & ECCV2022 & 621 & 203 & 38.17 / 0.9611 & 33.94 / 0.9207 & 32.30 / 0.9012 & 32.76 / 0.9340 & 39.12 / 0.9783\\
& SRFormer-light \cite{Zhou2023} & ICCV2023 & 853 & 236 & 38.23 / 0.9613 & 33.94 / 0.9209 & 32.36 / 0.9019 & 32.91 / 0.9353 & 39.28 / 0.9785\\
& MAN-light \cite{Wang2024b} & CVPRW2024 & 823 & 184 & 38.20 / 0.9613 & 33.95 / \textcolor{blue}{0.9214} & \textcolor{blue}{32.36} / \textcolor{blue}{0.9022} & \textcolor{blue}{32.92} / \textcolor{blue}{0.9364} & 39.40 / 0.9786\\
& MambaIR-light \cite{Guo2024} & ECCV2024 & 859 & 198.1 & 38.16 / 0.9610 & 34.00 / 0.9212 & 32.34 / 0.9017 & 32.92 / 0.9356 & 39.31 / 0.9779\\
& SeemoRe-L \cite{Zamfir2024} & ICML2024 & 953 & 202 & \textcolor{blue}{38.27} / \textcolor{blue}{0.9616} & \textcolor{blue}{34.01} / 0.9210 & 34.35 / 0.9018 & 32.87 / 0.9344 & \textcolor{blue}{39.49} / \textcolor{blue}{0.9790}\\
\rowcolor{gray!20}
\cellcolor{white!}& \textbf{LKFMixer-L} & Ours & 906 & 193 & \textcolor{red}{\textbf{38.31}} / \textcolor{red}{\textbf{0.9617}} & \textcolor{red}{\textbf{34.08}} / \textcolor{red}{\textbf{0.9223}} & \textcolor{red}{\textbf{32.42}} / \textcolor{red}{\textbf{0.9029}} & \textcolor{red}{\textbf{33.13}} / \textcolor{red}{\textbf{0.9371}} & \textcolor{red}{\textbf{39.57}} / \textcolor{red}{\textbf{0.9791}}\\
\midrule
\multirow{20}{*}{×3}
& PAN \cite{Zhao2020} & ECCV2020 & 261 & 39.1 & 34.10 / 0.9271 & 30.36 / 0.8423 & 29.11 / 0.8050 & 28.11 / 0.8511 & 33.61 / 0.9448\\
& SAFMN \cite{Sun2023} & ICCV2023 & 233 & 23 & 34.34 / 0.9267 & 30.33 / 0.8418 & 29.08 / 0.8048 & 27.95 / 0.8474 & 33.52 / 0.9437\\
& MAN-tiny \cite{Wang2024b} & CVPRW2024 & 141 & 14 & 34.24 / 0.9259 & 30.25 / 0.8405 & 29.04 / 0.8046 & 27.85 / 0.8465 & 33.28 / 0.9426\\
& SMFANet \cite{Zheng2024} & ECCV2024 & 191 & 19 & 34.42 / 0.9274 & 30.41 / 0.8430 & \textcolor{blue}{29.16} / \textcolor{blue}{0.8065} & 28.22 / 0.8523 & \textcolor{blue}{33.96} / \textcolor{blue}{0.9460}\\
& SeemoRe-T \cite{Zamfir2024} & ICML2024 & 225 & 21 & \textcolor{blue}{34.46} / \textcolor{blue}{0.9276} & \textcolor{blue}{30.44} / \textcolor{blue}{0.8445} & 29.15 / 0.8063 & \textcolor{blue}{28.27} / \textcolor{blue}{0.8538} & 33.92 / 0.9460\\
\rowcolor{gray!20}
\cellcolor{white!}& \textbf{LKFMixer-T} & Ours & 199 & 18.8 & \textcolor{red}{\textbf{34.51}} / \textcolor{red}{\textbf{0.9279}} & \textcolor{red}{\textbf{30.46}} / \textcolor{red}{\textbf{0.8447}} & \textcolor{red}{\textbf{29.18}} / \textcolor{red}{\textbf{0.8077}} & \textcolor{red}{\textbf{28.27}} / \textcolor{red}{\textbf{0.8541}} & \textcolor{red}{\textbf{33.97}} / \textcolor{red}{\textbf{0.9464}}\\
\cdashline{2-10}
& ShuffleMixer \cite{Sun2022} & NIPS2022 & 415 & 43 & 34.40 / 0.9272 & 30.37 / 0.8423 & 29.12 / 0.8051 & 28.08 / 0.8498 & 33.69 / 0.9448\\
& BSRN \cite{Li2022} & CVPRW2022 & 340 & 33 & 34.46 / 0.9277 & 30.47 / 0.8449 & 29.18 / 0.8068 & 28.39 / 0.8567 & 34.05 / 0.9471\\
& VapSR \cite{Zhou2022} & ECCV2022 & 351 & 35.6 & 34.52 / 0.9284 & \textcolor{blue}{30.53} / 0.8452 & 29.19 / 0.8077 & 28.43 / 0.8583 & 33.96 / 0.9469\\
& LKDN \cite{Xie2023} & CVPRW2023 & 311 & 31.6 & 34.54 / 0.9285 & 30.52 / \textcolor{blue}{0.8455} & 29.21 / 0.8078 & \textcolor{blue}{28.50} / \textcolor{blue}{0.8601} & \textcolor{blue}{34.08} / \textcolor{blue}{0.9475}\\
& CFSR \cite{Wu2024} & TIP2024 & 298 & 28.5 & 34.50 / 0.9279 & 30.44 / 0.8437 & 29.16 / 0.8066 & 28.29 / 0.8553 & 33.86 / 0.9462\\
& OSFFNet \cite{Wang2024} & AAAI2024 & 524 & 37.8 & \textcolor{blue}{34.58} / \textcolor{blue}{0.9287} & 30.48 / 0.8450 & \textcolor{blue}{29.21} / \textcolor{blue}{0.8080} & 28.49 / 0.8595 & 34.00 / 0.9472\\
\rowcolor{gray!20}
\cellcolor{white!}& \textbf{LKFMixer-B} & Ours & 364 & 34.4 & \textcolor{red}{\textbf{34.59}} / \textcolor{red}{\textbf{0.9287}} & \textcolor{red}{\textbf{30.57}} / \textcolor{red}{\textbf{0.8463}} & \textcolor{red}{\textbf{29.25}} / \textcolor{red}{\textbf{0.8095}} & \textcolor{red}{\textbf{28.58}} / \textcolor{red}{\textbf{0.8604}} & \textcolor{red}{\textbf{34.26}} / \textcolor{red}{\textbf{0.9481}}\\
\cdashline{2-10}
& SwinIR-light \cite{Liang2021} & ICCVW2021 & 918 & 114 & 34.62 / 0.9289 & 30.54 / 0.8463 & 29.20 / 0.8082 & 28.66 / 0.8624 & 33.98 / 0.9478\\
& ELAN-light \cite{Zhang2022} & ECCV2022 & 629 & 90.1 & 34.61 / 0.9288 & 30.55 / 0.8463 & 29.21 / 0.8081 & 28.69 / 0.8624 & 34.00 / 0.9478\\
& SRFormer-light \cite{Zhou2023} & ICCV2023 & 861 & 105 & 34.67 / 0.9296 & 30.57 / 0.8469 & 29.26 / 0.8099 & 28.81 / 0.8655 & 34.19 / 0.9489\\
& MAN-light \cite{Wang2024b} & CVPRW2024 & 832 & 82.7 & 34.66 / 0.9293 & 30.60 / \textcolor{blue}{0.8478} & \textcolor{blue}{29.29} / \textcolor{blue}{0.8105} & 28.87 / 0.8671 & 34.36 / 0.9492\\
& MambaIR-light \cite{Guo2024} & ECCV2024 & 867 & 88.7 & 34.72 / 0.9296 & \textcolor{blue}{30.63} / 0.8475 & 29.29 / 0.8099 & \textcolor{red}{29.00} / \textcolor{red}{0.8689} & 34.39 / 0.9495\\
& SeemoRe-L \cite{Zamfir2024} & ICML2024 & 959 & 90.5 & \textcolor{blue}{34.72} / \textcolor{blue}{0.9297} & 30.60 / 0.8469 & 29.29 / 0.8101 & 28.86 / 0.8653 & \textcolor{blue}{34.53} / \textcolor{blue}{0.9496}\\
\rowcolor{gray!20}
\cellcolor{white!}& \textbf{LKFMixer-L} & Ours & 915 & 86.8 & \textcolor{red}{\textbf{34.76}} / \textcolor{red}{\textbf{0.9301}} & \textcolor{red}{\textbf{30.66}} / \textcolor{red}{\textbf{0.8485}} & \textcolor{red}{\textbf{29.34}} / \textcolor{red}{\textbf{0.8118}} & \textcolor{blue}{\textbf{28.97}} / \textcolor{blue}{\textbf{0.8677}} & \textcolor{red}{\textbf{34.63}} / \textcolor{red}{\textbf{0.9503}}\\
\midrule
\multirow{21}{*}{×4}
& PAN \cite{Zhao2020} & ECCV2020 & 272 & 28.2 & 32.13 / 0.8948 & 28.61 / 0.7822 & 27.59 / 0.7363 & 26.11 / 0.7854 & 30.51 / 0.9095\\
& SAFMN \cite{Sun2023} & ICCV2023 & 240 & 14 & 32.18 / 0.8948 & 28.60 / 0.7813 & 27.58 / 0.7359 & 25.97 / 0.7809 & 30.43 / 0.9063\\
& MAN-Tiny \cite{Wang2024b} & CVPRW2024 & 150 & 8.4 & 32.07 / 0.8930 & 28.53 / 0.7801 & 27.51 / 0.7345 & 25.84 / 0.7786 & 30.18 / 0.9047\\
& SMFANet \cite{Zheng2024} & ECCV2024 & 197 & 11 & 32.25 / 0.8956 & 28.71 / 0.7833 & 27.64 / 0.7377 & 26.18 / 0.7862 & \textcolor{blue}{30.82} / \textcolor{blue}{0.9104}\\
& SeemoRe-T \cite{Zamfir2024} & ICML2024 & 232 & 11 & \textcolor{blue}{32.31} / \textcolor{blue}{0.8965} & \textcolor{blue}{28.72} / \textcolor{blue}{0.7840} & \textcolor{blue}{27.65} / \textcolor{blue}{0.7384} & \textcolor{blue}{26.23} / \textcolor{blue}{0.7883} & \textcolor{red}{30.82} / \textcolor{red}{0.9107}\\
\rowcolor{gray!20}
\cellcolor{white!}& \textbf{LKFMixer-T} & Ours & 207 & 11 & \textcolor{red}{\textbf{32.34}} / \textcolor{red}{\textbf{0.8965}} & \textcolor{red}{\textbf{28.74}} / \textcolor{red}{\textbf{0.7843}} & \textcolor{red}{\textbf{27.67}} / \textcolor{red}{\textbf{0.7391}} & \textcolor{red}{\textbf{26.23}} / \textcolor{red}{\textbf{0.7890}} & 30.76 / 0.9102\\
\cdashline{2-10}
& ShuffleMixer \cite{Sun2022} & NIPS2022 & 411 & 28 & 32.21 / 0.8953 & 28.66 / 0.7827 & 27.61 / 0.7366 & 26.08 / 0.7835 & 30.65 / 0.9093\\
& BSRN \cite{Li2022} & CVPRW2022 & 352 & 19.2 & 32.35 / 0.8966 & 28.73 / 0.7847 & 27.65 / 0.7387 & 26.27 / 0.7908 & 30.84 / 0.9123\\
& VapSR \cite{Zhou2022} & ECCV2022 & 342 & 19.8 & 32.38 / 0.8978 & 28.77 / 0.7852 & 27.68 / 0.7398 & 26.35 / 0.7941 & 30.86 / 0.9129\\
& LKDN \cite{Xie2023} & CVPRW2023 & 322 & 18.4 & \textcolor{blue}{32.39} / \textcolor{blue}{0.8979} & \textcolor{blue}{28.79} / \textcolor{blue}{0.7859} & \textcolor{blue}{27.69} / \textcolor{blue}{0.7402} & \textcolor{blue}{26.42} / \textcolor{red}{0.7965} & \textcolor{blue}{30.97} / \textcolor{blue}{0.9140}\\
& CFSR \cite{Wu2024} & TIP2024 & 307 & 17.5 & 32.33 / 0.8964 & 28.73 / 0.7842 & 27.63 / 0.7381 & 26.21 / 0.7897 & 30.72 / 0.9111\\
& OSFFNet \cite{Wang2024} & AAAI2024 & 537 & 22 & 32.39 / 0.8976 & 28.75 / 0.7852 & 27.66 / 0.7393 & 26.36 / 0.7950 & 30.84 / 0.9125\\
\rowcolor{gray!20}
\cellcolor{white!}& \textbf{LKFMixer-B} & Ours & 373 & 19.9 & \textcolor{red}{\textbf{32.46}} / \textcolor{red}{\textbf{0.8982}} & \textcolor{red}{\textbf{28.85}} / \textcolor{red}{\textbf{0.7865}} & \textcolor{red}{\textbf{27.75}} / \textcolor{red}{\textbf{0.7415}} & \textcolor{red}{\textbf{26.48}} / \textcolor{blue}{\textbf{0.7962}} & \textcolor{red}{\textbf{31.17}} / \textcolor{red}{\textbf{0.9148}}\\
\cdashline{2-10}
& SwinIR-light \cite{Liang2021} & ICCVW2021 & 930 & 65 & 32.44 / 0.8976 & 28.77 / 0.7858 & 27.69 / 0.7406 & 26.47 / 0.7980 & 30.92 / 0.9151\\
& ELAN-light \cite{Zhang2022} & ECCV2022 & 640 & 54.1 & 32.43 / 0.8975 & 28.78 / 0.7858 & 27.69 / 0.7406 & 26.54 / 0.7982 & 30.92 / 0.9150\\
& SRFormer-light \cite{Zhou2023} & ICCV2023 & 873 & 63 & 32.51 / 0.8988 & 28.82 / 0.7872 & 27.73 / 0.7422 & 26.67 / 0.8032 & 31.17 / 0.9165\\
& MAN-light \cite{Wang2024b} & CVPRW2024 & 840 & 47.1 & 32.50 / 0.8988 & 28.87 / 0.7885 & 27.77 / \textcolor{blue}{0.7429} & 26.70 / \textcolor{blue}{0.8052} & 31.25 / 0.9170\\
& CAMixerSR \cite{Wang2024a} & CVPR2024 & 765 & 77.9 & 32.51 / 0.8988 & 28.82 / 0.7870 & 27.72 / 0.7416 & 26.63 / 0.8012 & 31.18 / 0.9166\\
& MambaIR-Light \cite{Guo2024} & ECCV2024 & 879 & 50.6 & 32.51 / \textcolor{blue}{0.8993} & 28.85 / 0.7876 & 27.75 / 0.7423 & 26.75 / 0.8051 & 31.26 / 0.9175\\
& SeemoRe-L \cite{Zamfir2024} & ICML2024 & 969 & 51.4 & \textcolor{blue}{32.51} / 0.8990 & \textcolor{blue}{28.92} / \textcolor{blue}{0.7888} & \textcolor{blue}{27.78} / 0.7428 & \textcolor{blue}{26.79} / 0.8046 & \textcolor{blue}{31.48} / \textcolor{blue}{0.9181}\\
\rowcolor{gray!20}
\cellcolor{white!} & \textbf{LKFMixer-L} & Ours & 927 & 49.5 & \textcolor{red}{\textbf{32.71}} / \textcolor{red}{\textbf{0.9008}} & \textcolor{red}{\textbf{28.95}} / \textcolor{red}{\textbf{0.7892}} & \textcolor{red}{\textbf{27.83}} / \textcolor{red}{\textbf{0.7443}} & \textcolor{red}{\textbf{26.85}} / \textcolor{red}{\textbf{0.8069}} & \textcolor{red}{\textbf{31.52}} / \textcolor{red}{\textbf{0.9191}}\\
\bottomrule
\end{tabular}}
\end{table*}
\section{Experiments}
\label{Experiments}
\subsection{Datasets and indices}
\label{subsec:Datasets and indices}
Following previous works \cite{Wang2024,Zheng2024,Zamfir2024}, DIV2K \cite{Agustsson2017} and Flickr2K \cite{Lim2017} datasets are used for training. LR images are generated from HR images through bicubic downsampling. Five commonly used datasets-Set5 \cite{Bevilacqua2012}, Set14 \cite{Zeyde2012}, BSD100 \cite{Martin2001}, Urban100 \cite{Huang2015} and Manga109 \cite{Matsui2017} are adopted for testing. Meanwhile, PSNR and SSIM results are calculated on the Y-channel in the YCbCr color space.
\subsection{Implementation Details}
\label{subsec:Implementation Details}
The training HR images are decomposed into 480$\times$480 small pieces using sliding window slicing operation for faster training, with random rotation, horizontal and vertical flipping for data augment. The cropped LR patch size is 48$\times$48, and the batch size is 64. Similar to \cite{Zamfir2024}, we optimize $L_{1}$ loss and FFT loss using Adam \cite{Kingma2014} optimizer for 1000K iterations. The initial and minimum learning rates are set to 1$\times$$10^{-3}$ and 1$\times$$10^{-6}$, which are updated according to the Cosine Annealing scheme. We conduct all experiments with the PyTorch framework on an NVIDIA RTX 3090 GPU. The number of channels and FMBs are set to \{40, 48, 64\} and \{6, 8, 12\} for \{LKFMixer-T, LKFMixer-B, LKFMixer-L\}, respectively. For all LKFMixer models, the channel split factor $\alpha$ and large kernel size in PLKB are set to 0.25 and 31, respectively.
\subsection{Comparison with state-of-the-art methods}
\label{subsec:Comparison with state-of-the-art methods}
To evaluate the performance of the LKFMixer family, we compare it with several SOTA lightweight SR models, including PAN \cite{Zhao2020}, SAFMN \cite{Sun2023}, SMFANet \cite{Zheng2024}, SeeMoRe \cite{Zamfir2024}, ShuffleMixer \cite{Sun2022}, BSRN \cite{Li2022}, VapSR \cite{Zhou2022}, LKDN \cite{Wu2024}, CFSR \cite{Wu2024}, OSSFNet \cite{Wang2024}, SwinIR-light \cite{Liang2021}, ELAN-light \cite{Zhang2022}, SRFormer-light \cite{Zhou2023}, MAN \cite{Wang2024b}, MambaIR-light \cite{Guo2024}, and CAMixerSR-light \cite{Wang2024a}.
\subsubsection{Quantitative comparison}
\label{subsubsec:Quantitative comparison}
Table \ref{tab:Quantitative comparison results with other lightweight SR methods} presents the quantitative comparison results in terms of model parameters, computations, PSNR, and SSIM. As shown, our method almost achieves the best SR performance across all test datasets and scales. For instance, LKFMixer-T mostly outperforms SeeMoRe-T and achieves 0.08dB PSNR improvement on the Urban100 dataset at $\times$2 scale. LKFMixer-L shows remarkable superiority over several models, including SeeMoRe-L, CAMixerSR-light, and MambaIR-light, achieving 0.04dB, 0.34dB, and 0.26dB PSNR improvement on the Manga109 dataset at $\times$4 scale, respectively. These results demonstrate the effectiveness of the LKFMixer family across varying model complexities.
\subsubsection{Qualitative comparison}
\label{subsubsec:Qualitative comparison}
\begin{figure*}[!ht]
\centering
\includegraphics[width=1\linewidth]{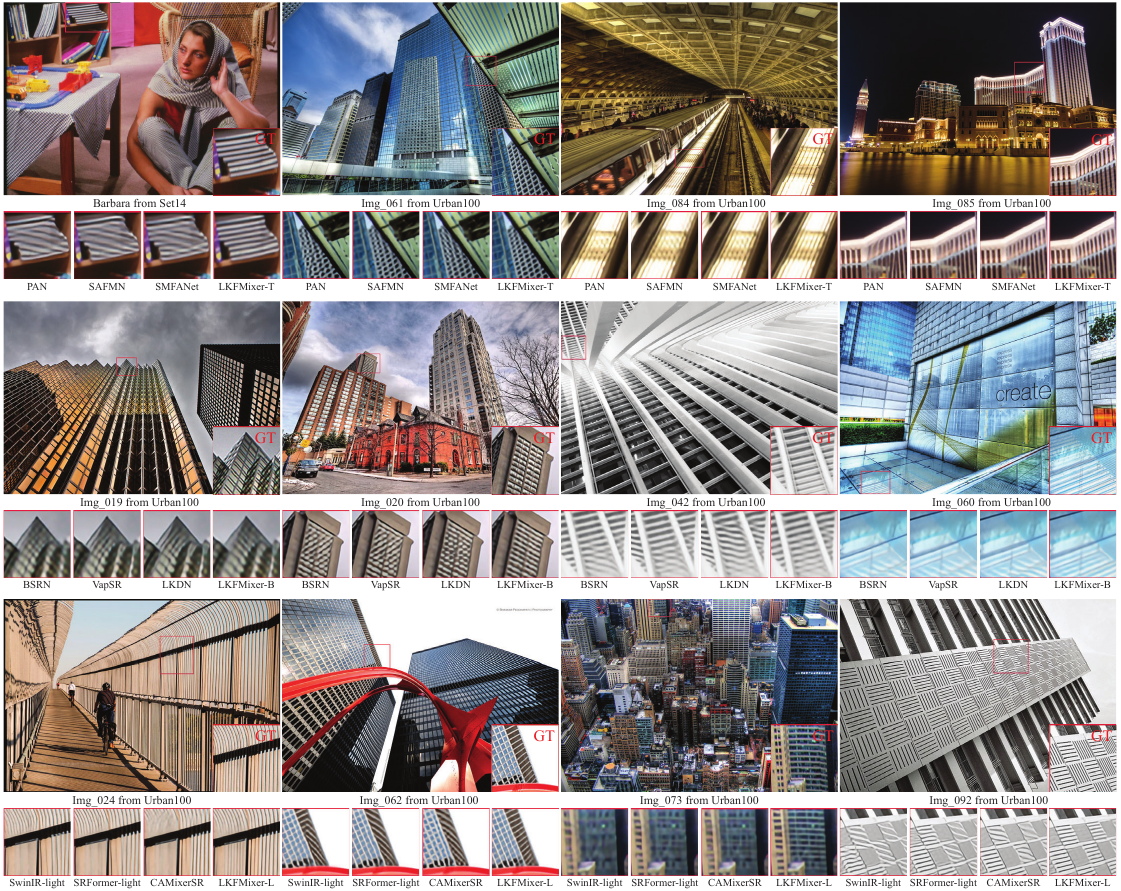}
\caption{Visual comparisons between LKFMixer family with other SOTA lightweight models for $\times$4 SR task. Zoom in for best view.}
\label{fig:Visual comparisons}
\end{figure*}

Fig. \ref{fig:Visual comparisons} illustrates the reconstruction results of LKFMixer family compared with other lightweight SR methods. As shown, our method consistently delivers superior reconstruction quality across three different model complexities. Specifically, only LKFMixer-T can recover the original structure of the image (e.g., Barbara from Set14), LKFMixer-B can recover more complete structural features with less distortion (e.g., Img-060 from Urban100), LKFMixer-L has a stronger reconstruction capability and can recover more texture details (e.g., Img-024 and Img-073 from Urban100). These results highlight the effectiveness of large kernel convolution for image reconstruction.
\subsubsection{LAM and ERF comparison}
\label{subsubsec:LAM and ERF comparison}
\begin{figure}[!ht]
\centering
\includegraphics[width=1\columnwidth]{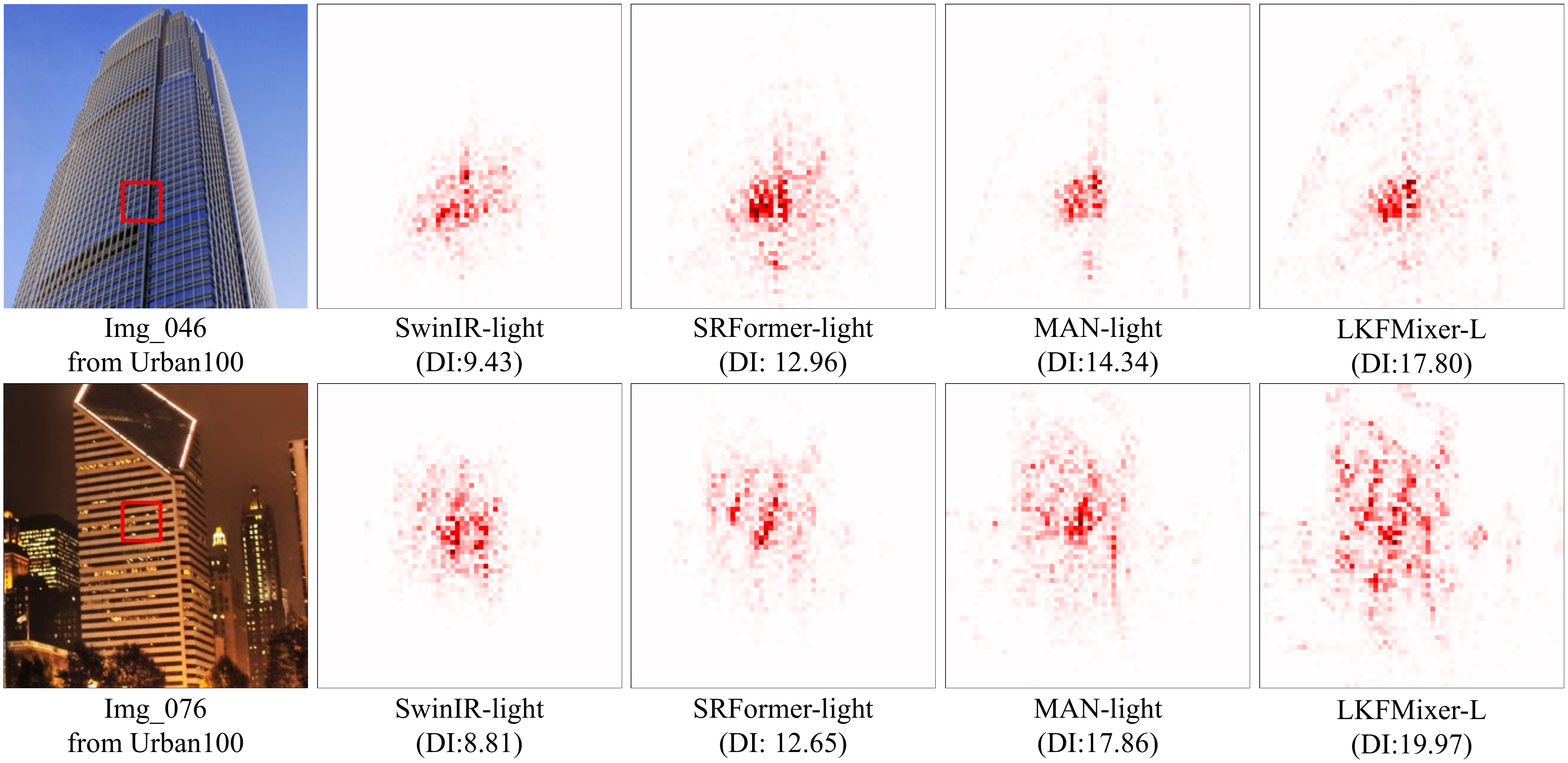}
\caption{Comparison of local attribution maps (LAMs) \cite{Gu2021} between LKFMixer-L with other lightweight SR models.}
\label{fig:Comparison results of LAM}
\end{figure}

\begin{figure}[!ht]
\centering
\includegraphics[width=1\columnwidth]{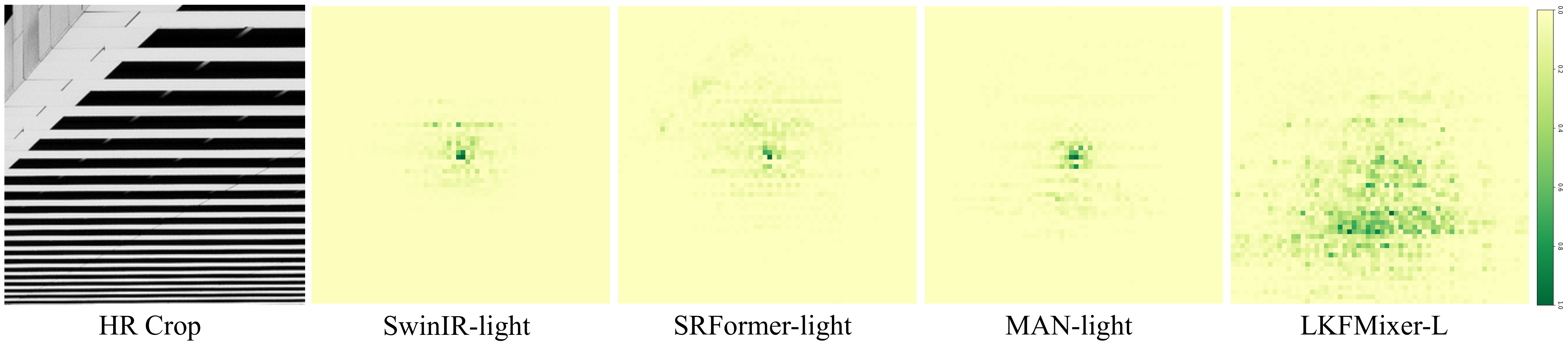}
\caption{Comparison of effective receptive field (EFR) \cite{Ding2022} between LKFMixer-L with other lightweight SR models.}
\label{fig:Comparison results of ERF}
\end{figure}

The local attribution map (LAM) \cite{Gu2021} is used to visualize the pixel range involved in image reconstruction, which can also be quantified using the diffusion index (DI). A larger DI indicates that more pixels are utilized for reconstruction. As shown in Fig. \ref{fig:Comparison results of LAM}, the DI value for LKFMixer-L is higher than that of other methods, suggesting that LKFMixer-L captures a larger number of pixels and thus has a broader receptive field. The effective receptive field (ERF) \cite{Ding2022}, which represents the actual receptive field that the model can access, is compared in Fig. \ref{fig:Comparison results of ERF}. The results indicate that LKFMixer-L has a larger ERF compared to the other models. These observations further validate the effectiveness of our large kernel convolution design in enhancing the receptive field.
\subsubsection{Memory and inference time comparison}
\label{subsubsec:Memory and inference time comparison}
To thoroughly verify the efficiency of the proposed model, we compare GPU memory consumption and inference time with other methods in Table \ref{tab:Inference time}. In terms of GPU memory consumption, the proposed model approximates other CNN-based models but is much lower than Transformer-based models (only 45\% of SwinIR-light). Regarding inference speed, the proposed model is equal to or slightly slower than CNN-based models but is much faster than Transformer-based models (almost $\times$5 faster than SwinIR-light). The results demonstrate the efficiency of the proposed model.
\begin{table}[!ht]
\caption{Comparison of average inference time and GPU memory consumption across 100 samples on 3090 GPU.}
\label{tab:Inference time}
\centering
\resizebox{\columnwidth}{!}{
\begin{tabular}{ccccc}
\toprule
Input & Scale & Method & \#GPU Mem.(M) & \#Avg. Time(ms)\\
\midrule
\multirow{11}{*}{[512, 512]} & \multirow{11}{*}{$\times4$} & PAN & 1196.09 & \textcolor{red}{51.72}\\
& & SeemoRe-T & \textcolor{red}{365.02} & 69.27\\
\rowcolor{gray!20}
\cellcolor{white!} & \cellcolor{white!} & \textbf{LKFMixer-T (Ours)} & \textcolor{blue}{\textbf{405.42}} & \textcolor{blue}{\textbf{61.02}}\\
\cdashline{3-5}
& & VapSR & \textcolor{red}{404.36} & 109.41\\
& & LKDN & 1176.29 & \textcolor{red}{68.93}\\
&  & \cellcolor{gray!20}\textbf{LKFMixer-B (Ours)} & \cellcolor{gray!20}\textcolor{blue}{\textbf{486.05}} & \cellcolor{gray!20}\textcolor{blue}{\textbf{98.51}}\\
\cdashline{3-5}
& & SwinIR-light & 1451.48 & 1000.89\\
& & SRFormer-light & 1303.83 & 998.50\\
& & CAMixerSR & 1489.17 & 311.72\\
& & SeemoRe-L & \textcolor{red}{470.94} & \textcolor{blue}{232.82}\\
\rowcolor{gray!20}
\cellcolor{white!} & \cellcolor{white!} & \textbf{LKFMixer-L (Ours)} & \textcolor{blue}{\textbf{648.21}} & \textcolor{red}{\textbf{194.43}}\\
\bottomrule
\end{tabular}
}
\end{table}
\begin{figure*}[!ht]
\centering
\includegraphics[width=1\linewidth]{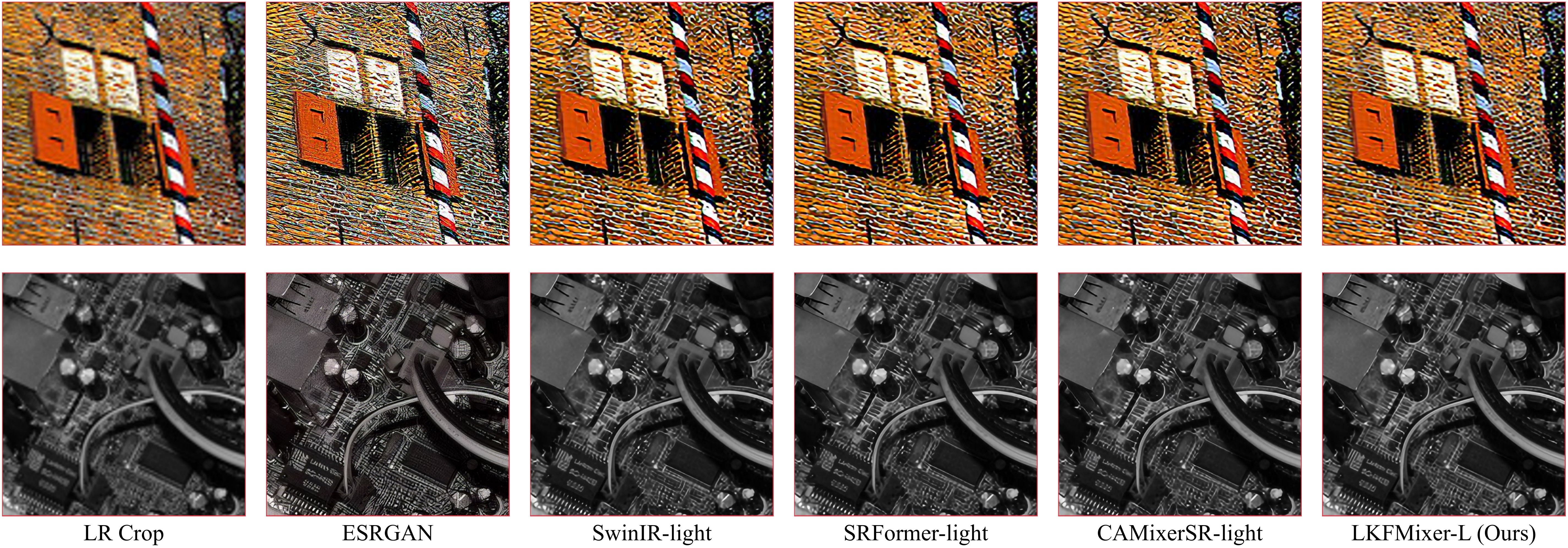}
\caption{Visual comparison with ESRGAN \cite{Wang2018a}, SwinIR-light \cite{Liang2021}, SRFormer-light \cite{Zhou2023}, CAMixerSR-light \cite{Wang2024a}, and the proposed LKFMixer-L on \textbf{real-word images} at $\times4$ upscale.}
\label{fig:Visual comparison on real-word images}
\end{figure*}
\begin{table*}[!ht]
\caption{Ablation study based on LKFMixer-B at $\times$4 scale. Metrics are also measured on the Y-channel. ``A$\rightarrow$B'' is the operation of replacing A with B, and ``A$\rightarrow$None'' is the operation of removing A. ``-'' denotes remain unchanged. WSA represents the window-based self-attention used in SwinIR \cite{Liang2021}.}
\label{tab:Ablation study}
\centering
\resizebox{\linewidth}{!}{
\begin{tabular}{clcccccccccc}
\toprule
\multirow{2}{*}{Ablation} & \multirow{2}{*}{Variant} & \#Params & \#Flops & \#GPU Mem. & \#Time & Set5 & Set14 & BSD100 & Urban100 & Manga109 & DIV2K100 \\
& & [K] & [G] & [M] & [ms] & PSNR / SSIM & PSNR / SSIM & PSNR / SSIM & PSNR / SSIM & PSNR / SSIM & PSNR / SSIM\\
\midrule
\rowcolor{gray!20}
Baseline &  LKFMixer-B & 373 & 19.9 & 486.05 & 98.51 & 32.46 / 0.8982 & 28.85 / 0.7865 & 27.75 / 0.7415 & 26.48 / 0.7962 & 31.17 / 0.9148 & 30.71 / 0.8432\\
\midrule
\multirow{3}{*}{FMB} &  FDB $\rightarrow$ None & 155 & 7.6 & 483.63 & 40.91 & 32.09 / 0.8936 & 28.57 / 0.7804 & 27.57 / 0.7360 & 25.90 / 0.7783 & 30.27 / 0.9040 & 30.42 / 0.8371\\
&  SFMB $\rightarrow$ None & 306 & 17.3 & 485.77 & 81.00 & 32.43 / 0.8979 & 28.81 / 0.7860 & 27.72 / 0.7409 & 26.45 / 0.7949 & 30.99 / 0.9126 & 30.61 / 0.8420\\
&  FSB $\rightarrow$ None & 307 & 16.1 & 485.78 & 74.52 & 32.40 / 0.8978 & 28.80 / 0.7859 & 27.73 / 0.7408 & 26.39 / 0.7935 & 31.03 / 0.9127 & 30.66 / 0.8425\\
\midrule
\multirow{2}{*}{PLKB} & 1$\times$31+31$\times$1 $\rightarrow$ 31$\times$31 & 804 & 44.7 & 487.69 & 194.79 & 32.45 / 0.8978 & 28.82 / 0.7862 & 27.74 / 0.7414 & 26.48 / 0.7964 & 31.09 / 0.9143 & 30.69 / 0.8432\\
& PLKB $\rightarrow$ WSA & 678 & 49.3 & 1490.03 & 1354.31 & 32.52 / 0.8987 & 28.89 / 0.7880 & 27.77 / 0.7424 & 26.59 / 0.7996 & 31.32 / 0.9169 & 30.74 / 0.8440\\
\midrule
\multirow{2}{*}{FFB} &  DWConv3 $\rightarrow$ None & 361 & 19.3 & 485.99 & 85.99 & 32.42 / 0.8975 & 28.82 / 0.7861 & 27.74 / 0.7413 & 26.48 / 0.7965 & 31.11 / 0.9141 & 30.69 / 0.8429\\
&  PLKB $\rightarrow$ None & 298 & 15.6 & 485.73 & 67.81 & 32.36 / 0.8974 & 28.77 / 0.7854 & 27.72 / 0.7404 & 26.36 / 0.7928 & 30.98 / 0.9129 & 30.64 / 0.8420\\
\midrule
\multirow{6}{*}{SFMB} &  PLKB $\rightarrow$ None & 348 & 18.5 & 485.94 & 86.95 & 30.61 / 0.8609 & 27.41 / 0.7453 & 26.88 / 0.7103 & 24.31 / 0.7085 & 26.72 / 0.8131 & 29.08 / 0.7983\\
& Channel attention $\rightarrow$ None & 373 & 19.8 & 486.05 & 96.65 & 32.45 / 0.8981 & 28.83 / 0.7864 & 27.74 / 0.7415 & 26.51 / 0.7969 & 31.06 / 0.9143 & 30.70 / 0.8433\\
& Spatial branch $\rightarrow$ None & 350 & 19.7 & 485.95 & 92.14 & 32.35 / 0.8970 & 28.79 / 0.7859 & 27.72 / 0.7406 & 26.36 / 0.7929 & 30.96 / 0.9127 & 30.67 / 0.8425\\
& Downsampling factor 8 $\rightarrow$ 4 & 373 & 19.9 & 486.05 & 99.46 & 32.42 / 0.8977 & 28.81 / 0.7862 & 27.73 / 0.7410 & 26.45 / 0.7953 & 31.11 / 0.9140 & 30.68 / 0.8424\\
& Downsampling factor 8 $\rightarrow$ 2 & 373 & 20.2 & 486.05 & 101.96 & 32.46 / 0.8979 & 28.84 / 0.7861 & 27.74 / 0.7411 & 26.49 / 0.7964 & 31.02 / 0.9140 & 30.69 / 0.8428\\
& Addition $\rightarrow$ Concatenetion & 391 & 20.9 & 486.12 & 99.85 & 32.43 / 0.8977 & 28.84 / 0.7868 & 27.75 / 0.7417 & 26.48 / 0.7963 & 31.09 / 0.9144 & 30.69 / 0.8433\\
\midrule
\multirow{2}{*}{FSB} & Concatenation$\rightarrow$Addition & 354 & 18.7 & 485.98 & 97.19 & 32.45 / 0.8980 & 28.81 / 0.7862 & 27.73 / 0.7410 & 26.45 / 0.7961 & 31.08 / 0.9141 & 30.69 / 0.8428\\
& 1-Sigmoid$\rightarrow$Sigmoid & - & - & - & - & 32.45 / 0.8981 & 28.82 / 0.7861 & 27.73 / 0.7413 & 26.46 / 0.7957 & 31.10 / 0.9139 & 30.69 / 0.8431\\
\midrule
\multirow{2}{*}{Loss function} & L1 loss $\rightarrow$ None & - & - & - & - & 32.42 / 0.8980 & 28.82 / 0.7863 & 27.73 / 0.7410 & 26.49 / 0.7962 & 31.10 / 0.9143 & 30.70 / 0.8429\\
& FFT loss $\rightarrow$ None & - & - & - & - & 32.39 / 0.8970 & 28.78 / 0.7857 & 27.71 / 0.7406 & 26.39 / 0.7964 & 30.85 / 0.9125 & 30.64 / 0.8427\\
\bottomrule
\end{tabular}
}
\end{table*}
\subsubsection{Visual perception comparison}
\label{subsubsec:Visual perception comparison}
To further prove the effectiveness of LKFMixer, we adopt perceptual metric LPIPS \cite{Zhang2018a} to evaluate the visual quality of the reconstructed images, and the results are shown in Table \ref{tab:LPIPS comparison}. The lower the LPIPS is, the more realistic the recovered image is. Compared with other lightweight Transformer-based methods, our method almost achieves the lowest LPIPS values on all three datasets, except for slightly higher than SRFormer-light \cite{Zhou2023} on the Urban100 dataset. This proves that the images recovered by the proposed LKFMixer exhibit superior visual perception and quality.
\begin{table}[!ht]
\caption{LPIPS \cite{Zhang2018a} comparison on several benchmark datasets. The lower the LPIPS, the better the visual perception of images.}
\label{tab:LPIPS comparison}
\centering
\resizebox{\columnwidth}{!}{
\begin{tabular}{ccccc}
\toprule
Datasets & SwinIR-light \cite{Liang2021} & SRFormer-light \cite{Zhou2023} & CAMixerSR-light \cite{Wang2024a} & LKFMixer-L (Ours)\\
\midrule
Urban100 & 0.2161 & \textbf{0.2109} & 0.2150 & \underline{0.2123} \\
Manga109 & 0.1021 & \underline{0.1005} & 0.1009 & \textbf{0.1001} \\
DIV2K100 & 0.2634 & 0.2608 & \underline{0.2595} & \textbf{0.2595} \\
\bottomrule
\end{tabular}
}
\end{table}
\par
Meanwhile, to verify the generalization of the proposed model, we compare LKFMixer-L with other SR methods in real-world images, and the results are shown in Fig. \ref{fig:Visual comparison on real-word images}. ESRGAN \cite{Wang2018a} excels in recovering more realistic images by using perceptual loss. However, our LKFMixer-L and other Transformer-based SR models, tends to produce smoother images while still effectively recovering the overall structure and texture details. These results highlight that LKFMixer-L has a robust ability to recover image details, even when processing real-world images with potentially unknown or complex degradation patterns.
\section{Analysis and Discussion}
\label{sec:Analysis and Discussion}
In this section, we conduct extensive ablation experiments to evaluate the effectiveness of the internal modules and large kernel design, as shown in Table \ref{tab:Ablation study}. For the three modules FDB, SFMB and FSB, the model performance decreases when one of them is individually removed, indicating the effectiveness of each module. Next, we provide a detailed analysis of their internal design.
\subsection{Effectiveness of the PLKB}
\label{subsec:Effectiveness of the PLKB}
Considering that the large convolution kernel is the core of PLKB, we first replace the serial strip convolutions with DWConv31$\times$31, and the model performance remains unchanged or even decreases. However, the model parameters increase by 115.5\%, and the inference time increases by 97.7\%, which proves that strip convolution is more suitable for lightweight models. To investigate the impact of kernel size on model performance, we gradually increase the kernel size from 3 to 41, and the results are shown in Table \ref{tab:Ablation study of large kernel size}. The overall model performance gradually improves and peaks when the kernel size reaches 31, but the model performance declines when the kernel size is further increased to 41. Importantly, the increase in kernel size does not lead to a significant rise in model parameters and inference time, thanks to the coordinate decomposition and partial channel design.
\par
\begin{table}[!ht]
\caption{Ablation study of large kernel size in PLKB. The large kernel form is 1$\times$K and K$\times$1.}
\label{tab:Ablation study of large kernel size}
\centering
\resizebox{\columnwidth}{!}{
\begin{tabular}{ccccccccc}
\toprule
\multirow{2}{*}{Kernel} & \#Params & \#Flops & \#GPU Mem. & \#Time & BSD100 & Urban100 & Manga109 & DIV2K100\\
& [K] & [G] & [M] & [ms] & PSNR / SSIM & PSNR / SSIM & PSNR / SSIM & PSNR / SSIM\\
\midrule
3   & 346 & 18.3 & 485.97 & 82.35 & 27.71 / 0.7403 & 26.40 / 0.7935 & 31.05 / 0.9135 & 30.66 / 0.8421\\
11 & 354 & 18.8 & 486.01 & 86.71 & 27.72 / 0.7405 & 26.44 / 0.7948 & 31.08 / 0.9139 & 30.68 / 0.8426\\
21 & 363 & 19.3 & 486.01 & 91.64 & 27.72 / 0.7407 & 26.45 / 0.7953 & 31.11 / 0.9144 & 30.69 / 0.8430\\
\rowcolor{gray!20}
31 & 373 & 19.9 & 486.05 & 98.51 & 27.75 / 0.7415 & 26.48 / 0.7962 & 31.17 / 0.9148 & 30.71 / 0.8432\\
41 & 382 & 20.4 & 486.09 & 103.81 & 27.73 / 0.7413 & 26.45 / 0.7953 & 30.92 / 0.9129 & 30.66 / 0.8426\\
\bottomrule
\end{tabular}
}
\end{table}
\par
\begin{figure}[!ht]
\centering
\includegraphics[width=1\columnwidth]{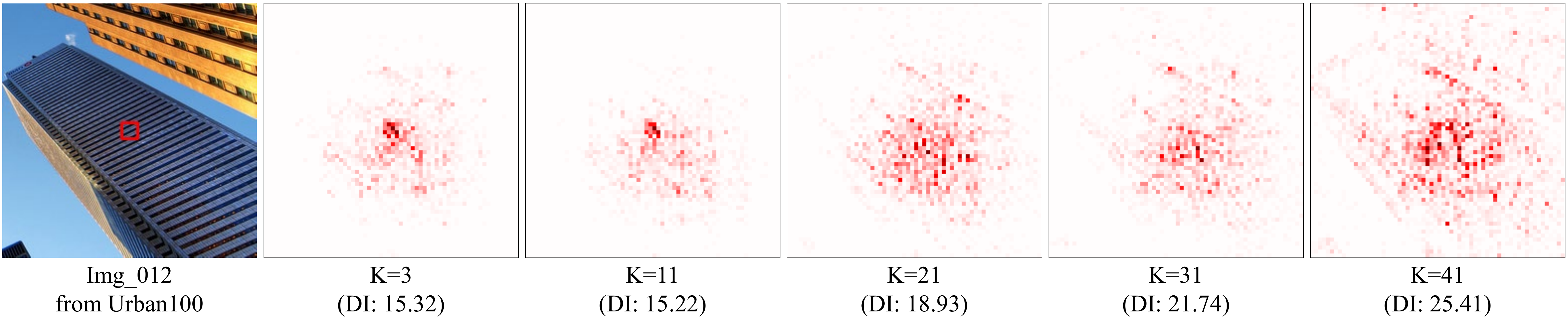}
\caption{LAM comparison of different kernel size in PLKB.}
\label{fig:LAM comparison results of kernel in PLKB}
\end{figure}
\begin{figure}[!ht]
\centering
\includegraphics[width=1\columnwidth]{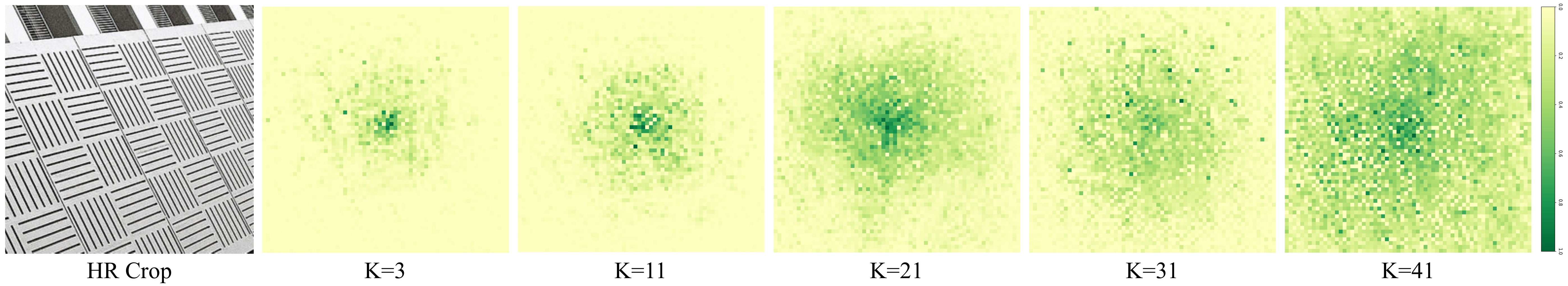}
\caption{ERF analysis of different kernel size in PLKB.}
\label{fig:ERF comparison results of kernel in PLKB}
\end{figure}

We also analyze the relationship between kernel size and receptive field utilizing LAM and ERF, as shown in Fig. \ref{fig:LAM comparison results of kernel in PLKB} and Fig. \ref{fig:ERF comparison results of kernel in PLKB}, respectively. It is evident that the receptive field increases as the kernel size grows and reaches maximum when the kernel size is 41. Taking the model performance into consideration, we ultimately select the kernel size of 31 for the PLKB. Additionally, we replace PLKB with the WSA (window-based self-attention) mechanism \cite{Liang2021}, which leads to a significant improvement in model performance. However, this comes at the cost of 81.8\% increase in model parameters, 206.6\% increase in GPU memory consumption, and notably $\times$13 times increase in inference time.
\par
\begin{figure}[!ht]
\centering
\includegraphics[width=1\linewidth]{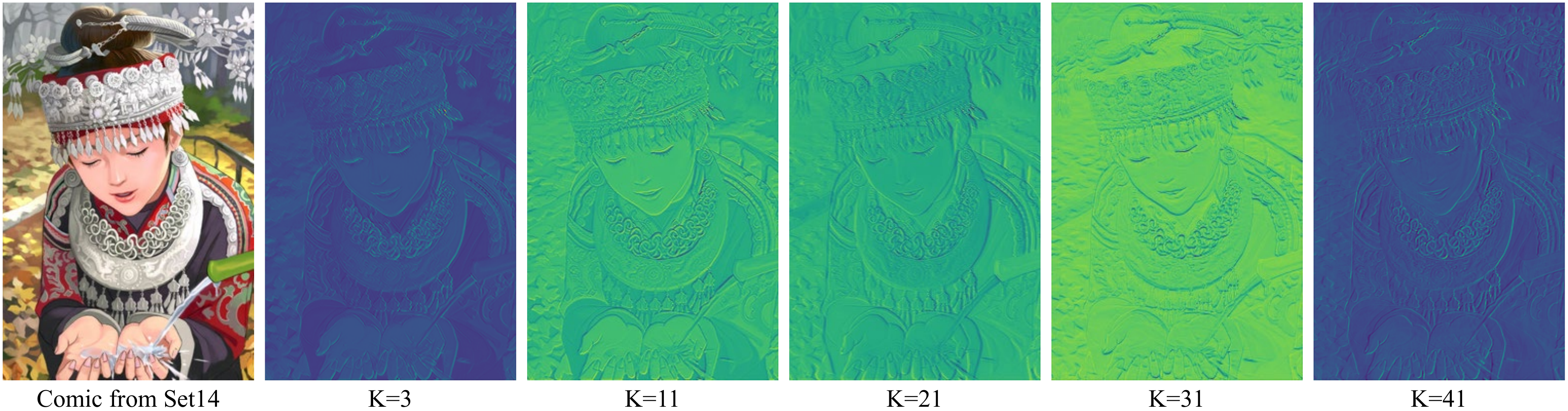}
\caption{Feature map visualization of PLKB output with different kernel sizes.}
\label{fig:Feature map visualization of PLKB output with different kernel sizes}
\end{figure}

We further analyze how the size of the convolution kernel in PLKB impacts the extracted feature maps, and the results are shown in Fig. \ref{fig:Feature map visualization of PLKB output with different kernel sizes}. When the convolution kernel size is 3, the output feature map primarily contains high-frequency details such as edges and textures. This highlights the small convolution kernel's strong ability to capture local features. As the kernel size increases to 31, the feature map begins to include more low-frequency information, such as image contours, while still retaining high-frequency details. This demonstrates the large convolution kernel's ability to capture non-local features. However, when the kernel size is increased further to 41, the high-frequency details become more pronounced, while the low-frequency contour information diminishes. The change in the feature map is consistent with the observed variation in SR performance as the kernel size increases, further validating the effectiveness of larger convolution kernels in enhancing image reconstruction capabilities.
\par
\begin{figure}[!ht]
\centering
\includegraphics[width=1\columnwidth]{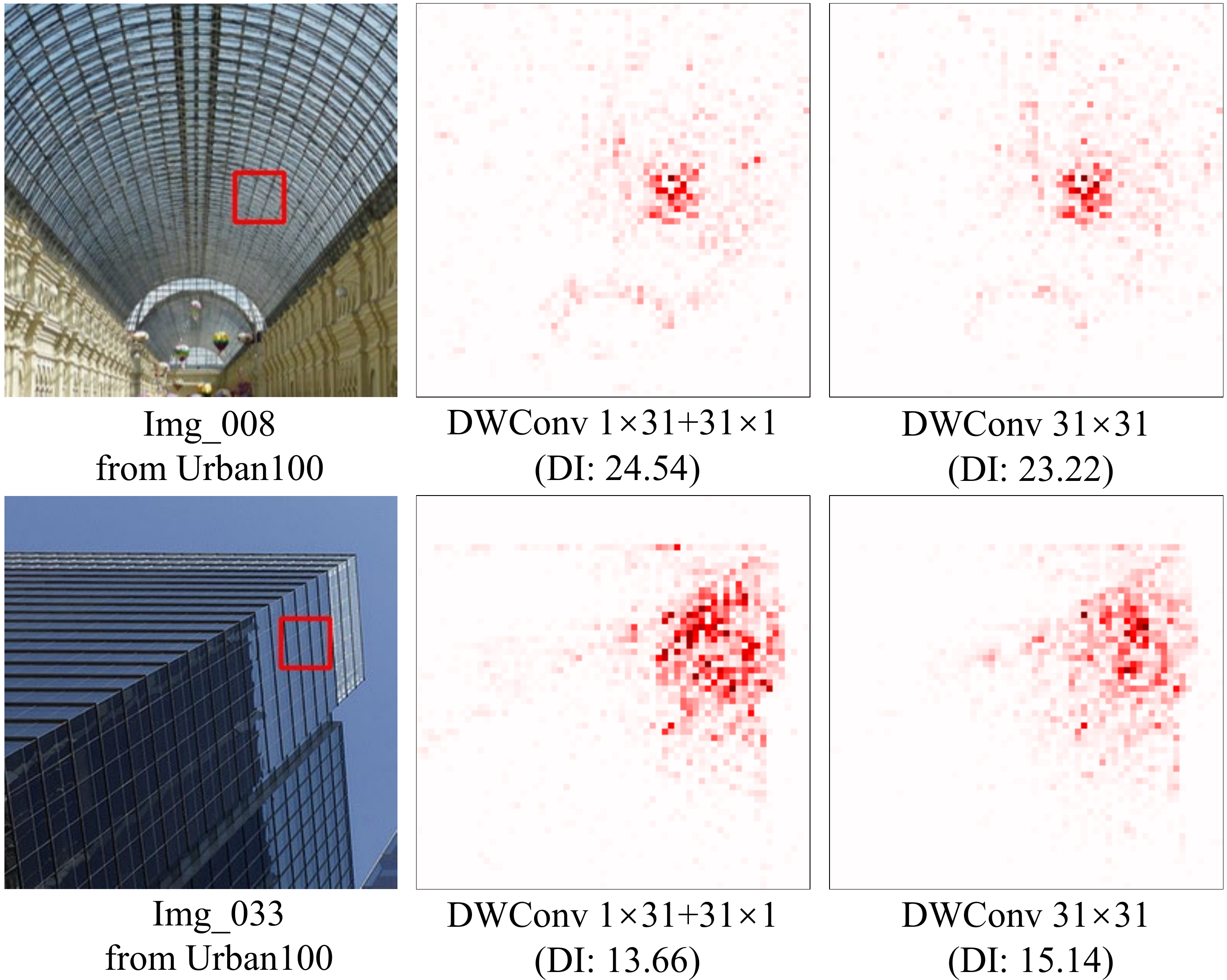}
\caption{LAM comparison between large kernel DWConv31$\times$31 and decomposed series strip large kernel DWConv1$\times$31+31$\times$1 at $\times4$ upscale.}
\label{fig:appendix_LAM comparison on large kernel}
\end{figure}

To distinguish the receptive field differences between DWConv31$\times$31 and the decomposed DWConv1$\times$31+31$\times$1, the LAM \cite{Gu2021} results of them are shown in Fig. \ref{fig:appendix_LAM comparison on large kernel}. The LAM range for both convolutions is nearly identical, suggesting that the structure of the convolution kernel has minimal impact on the receptive field as long as the kernel size remains unchanged.
\par
\begin{figure}[!ht]
\centering
\includegraphics[width=1\linewidth]{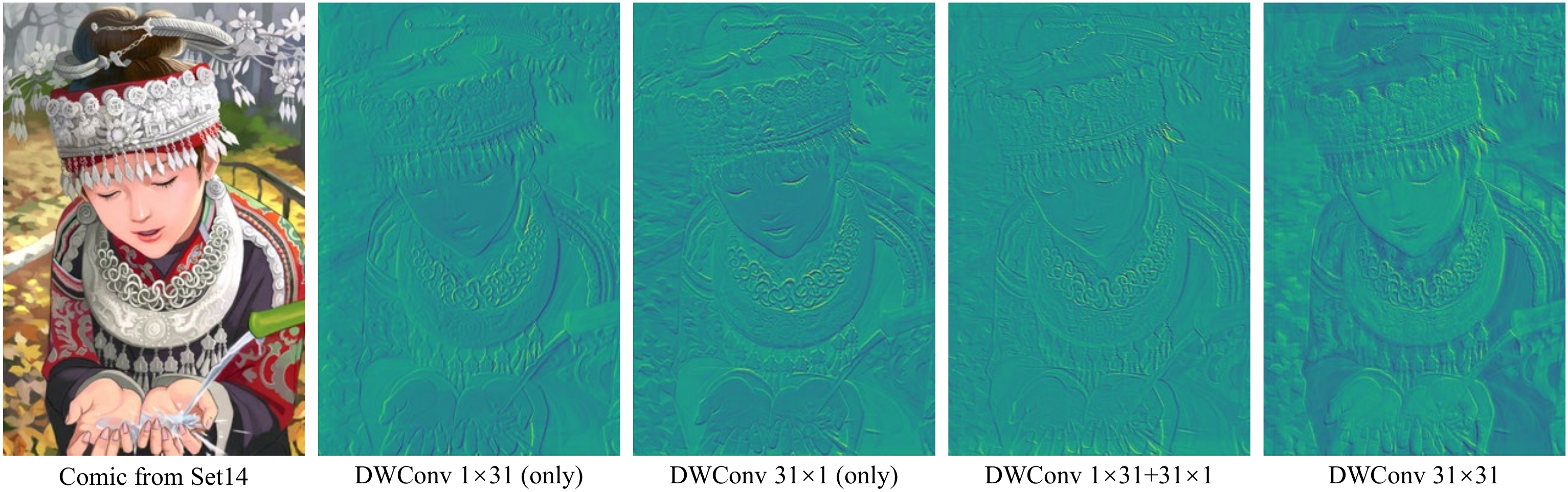}
\caption{Feature map visualization of different large kernel output in PLKB.}
\label{fig:Feature map visualization of large kernel output in PLKB}
\end{figure}

In addition, we investigate the differences in feature maps extracted by the two convolution kernel structures, and the results are shown in Fig. \ref{fig:Feature map visualization of large kernel output in PLKB}. The features extracted by DWConv1$\times$31 and DWConv31$\times$1 are almost identical. However, slight differences can be observed in the details (such as leaf and clothing textures) due to the strip structure and operation logic of the convolution kernels. Since DWConv1$\times$31+31$\times$1 consists of two serial strip convolutions, the features it extracts synthesize those of DWConv1$\times$31 and DWConv31$\times$1, with rich texture details while preserving overall contour features. On the other hand, DWConv31$\times$31 can also extract low-frequency contour information and high-frequency edge or texture features. Considering the balance between model complexity and inference speed, we conclude that the decomposed strip large convolution kernel is better suited for the lightweight SR model.
\par
\begin{figure}[!ht]
\centering
\includegraphics[width=1\columnwidth]{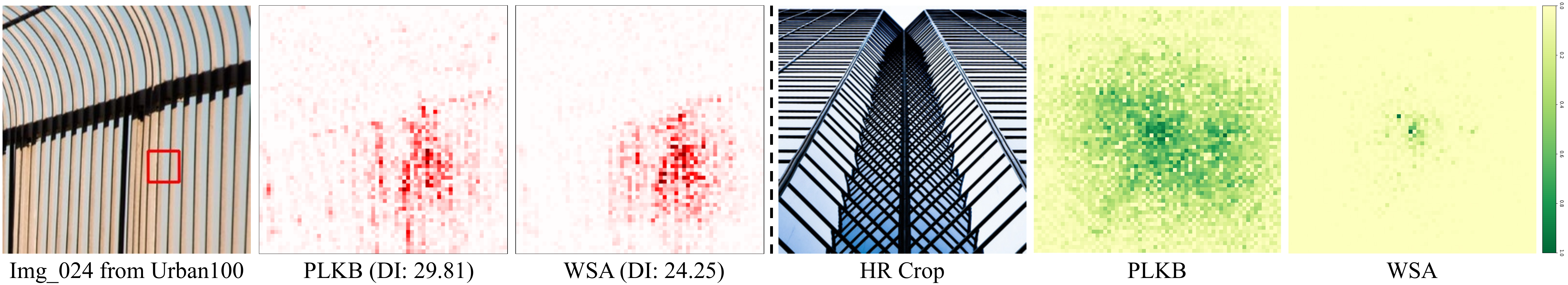}
\caption{LAM (left) and ERF (right) comparison of PLKB and WSA \cite{Liang2021}.}
\label{fig:LAM and ERF comparison of PLKB and WSA}
\end{figure}

Meanwhile, we utilize LAM and ERF to analyze the receptive field of PLKB and WSA, as shown in Fig. \ref{fig:LAM and ERF comparison of PLKB and WSA}. The results demonstrate that PLKB achieves a larger receptive field, indicating that it can capture non-local features similar to WSA in Transformer-based models, but is more suitable for lightweight SR models.
\par
Furthermore, we analyze the influence of channel split factor $\alpha$ on model performance, as shown in Table \ref{tab:Ablation study of channel split factor}. As $\alpha$ increase, more channels participate in the convolution operation, leading to model parameters and inference time gradually increase, but the model performance does not increase significantly. After comprehensive consideration, we select $\alpha$=0.25 to maintain an optimal balance between model parameters, SR performance, and inference time.
\begin{table}[!ht]
\caption{Ablation study of channel split factor $\alpha$ in PLKB.}
\label{tab:Ablation study of channel split factor}
\centering
\resizebox{\columnwidth}{!}{
\begin{tabular}{ccccccccc}
\toprule
\multirow{2}{*}{$\alpha$} & \#Params & \#Flops & \#GPU Mem. & \#Time & BSD100 & Urban100 & Manga109 & DIV2K100\\
& [K] & [G] & [M] & [ms] & PSNR / SSIM & PSNR / SSIM & PSNR / SSIM & PSNR / SSIM\\
\midrule
0.125 & 357 & 19.0 & 486.01 & 88.24 & 27.73 / 0.7410 & 26.48 / 0.7956 & 31.10 / 0.9142 & 30.68 / 0.8427\\
\rowcolor{gray!20}
0.25 & 373 & 19.9 & 486.05 & 98.51 & 27.75 / 0.7415 & 26.48 / 0.7962 & 31.17 / 0.9148 & 30.71 / 0.8432\\
0.5   & 404 & 21.6 & 486.17 & 116.13 & 27.74 / 0.7415 & 26.49 / 0.7965 & 31.15 / 0.9148 & 30.71 / 0.8433\\
0.75 & 434 & 23.3 & 496.76 & 136.20 & 27.74 / 0.7420 & 26.54 / 0.7990 & 30.98 / 0.9134 & 30.72 / 0.8438\\
1      & 465 & 25.0 & 508.88 & 156.44 & 27.76 / 0.7419 & 26.48 / 0.7966 & 31.12 / 0.9147 & 30.72 / 0.8438\\
\bottomrule
\end{tabular}
}
\end{table}
\par
\begin{figure}[!ht]
\centering
\includegraphics[width=1\linewidth]{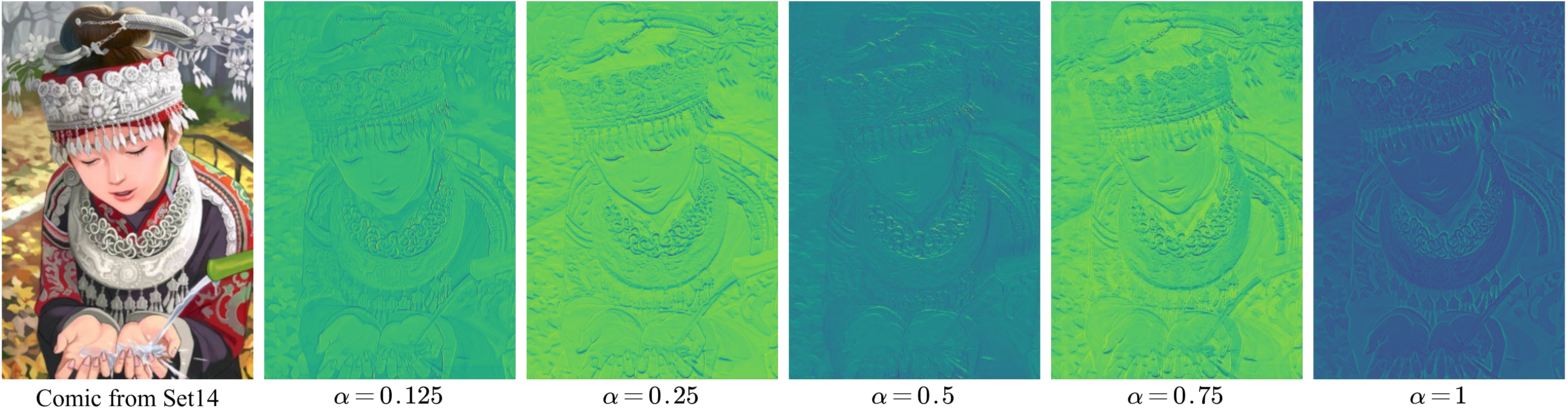}
\caption{Feature map visualization of PLKB output with different channel split factor $\alpha$.}
\label{fig:Feature map visualization of PLKB output with different channel split factor}
\end{figure}

The experimental results in Table \ref{tab:Ablation study of channel split factor} reveal that as the number of channels increases, the overall model performance improves. To gain a deeper understanding of this relationship, we further analyze the feature map outputs of PLKB under different channel split factors $\alpha$, and the results are shown in the Fig. \ref{fig:Feature map visualization of PLKB output with different channel split factor}. As $\alpha$ increases, more channels are employed for feature extraction, resulting in a feature map that contains more texture details while retaining the essential contour information, which is beneficial for recovering more accurate image details. However, considering both the model complexity and inference time, we determine that $\alpha$=0.25 offers an optimal balance between performance and efficiency, allowing us to maintain the ability to capture detailed features without excessively increasing model complexity.
\subsection{Effectiveness of the FFB}
\label{subsec:Effectiveness of the FFB}
\begin{figure}[!ht]
\centering
\includegraphics[width=1\columnwidth]{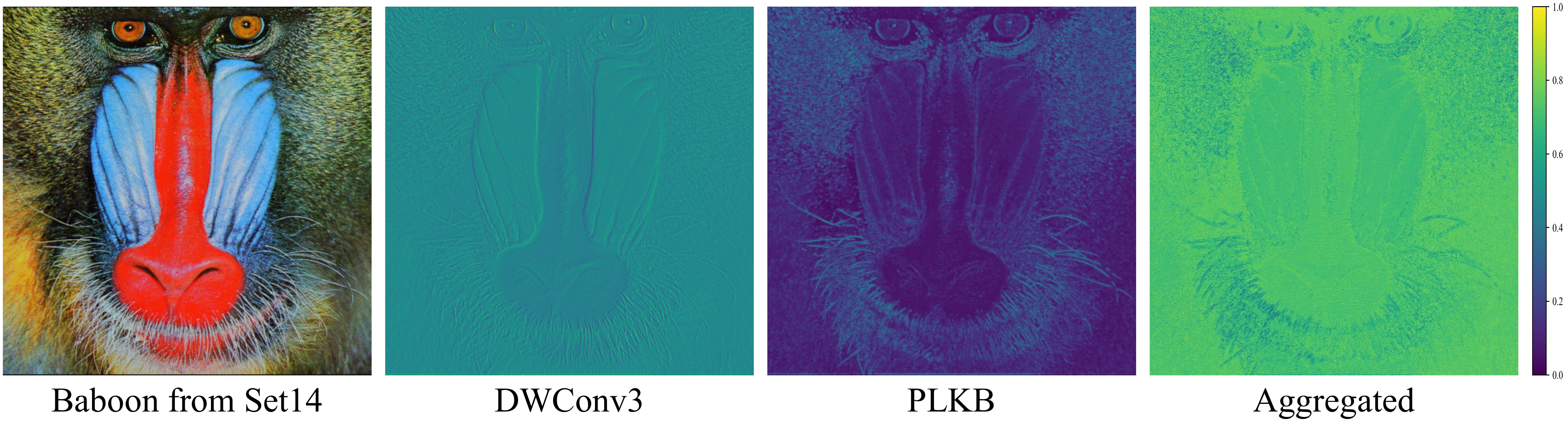}
\caption{Visualization of different feature map outputs from FFB.}
\label{fig:Visualization of feature maps}
\end{figure}

PLKB and DWConv3$\times$3 are responsible for capturing non-local and local features, respectively. Fusing the outputs of them is beneficial for obtaining more comprehensive feature information. When either PLKB or DWConv3$\times$3 is individually removed, the model performance decreases, indicating that the features from them are complementary. To further analyze the feature differences they extract, we visualize their feature maps in Fig. \ref{fig:Visualization of feature maps}. It can be observed that DWConv3$\times$3 can extract local details (e.g., whiskers), while PLKB can further capture structural features (e.g., contours). Both local and global features can be effectively preserved after aggregating them.
\par
\begin{figure}[!ht]
\centering
\includegraphics[width=1\columnwidth]{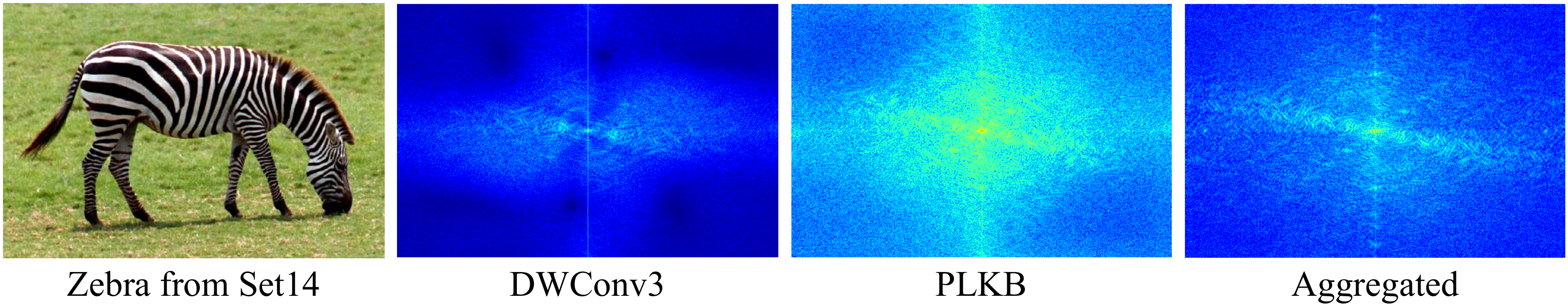}
\caption{Visualization of power spectral density (PSD) outputs from FFB, which has shifted the low-frequency component of the spectrum to the center by periodic displacement.}
\label{fig:Visualization of PSD}
\end{figure}

Following \cite{Zheng2024}, we also utilize power spectral density (PSD) to analyze the frequency-domain features, as shown in Fig. \ref{fig:Visualization of PSD}. The results indicate that DWConv3$\times$3 tends to capture sparse high-frequency features, while PLKB simultaneously acquires rich high-frequency and low-frequency features in the center of the spectrum. Meanwhile, both low-frequency and high-frequency features can be retained after information aggregation.
\subsection{Effectiveness of the SFMB}
\label{subsec:Effectiveness of the SFMB}
For both PLKB and spatial branches, the model performance decreases significantly when either of them is removed, indicating that non-local features and low-frequency spatial information are essential for SR reconstruction. Similarly, when we remove the channel attention in the spatial branch, the overall model performance decreases, emphasizing the importance of focusing on the salient features in channels. Regarding the downsampling factor in the spatial branch, the model performance drops when we reduce the downsampling factor from 8 to 4 or 2. The reason may be that the abstract information extracted by larger downsampling scale is more helpful to the SR model. Additionally, when we change the fusion mode of the large kernel branch and spatial branch from element-wise addition to channel concatenation, the model complexity increases and the overall model performance decreases. So we utilize element-wise addition for information fusion in SFMB to maintain efficiency and performance.
\subsection{Effectiveness of the SFB}
\label{subsec:Effectiveness of the SFB}
We first analyze the processing strategy of the sigmoid function. If the weight of PLKB branch changes from 1-$\beta$ to $\beta$ while the DWConv3$\times$3 branch is also weighted by $\beta$, we observe a degradation in model performance, suggesting that maintaining complementary weights between the branches is beneficial for improving model performance. In addition, we analyze the feature aggregation mode of PLKB and DWConv3$\times$3 features. When the channel concatenation operation is replaced with the element-wise addition, the model performance decreases although the model parameters are slightly reduced, which indicates that the information fusion strategy of channel concatenation is more effective for SFB.
\section{Limitations and Future Work}
\label{sec:Limitations and Future Work}
In this paper, we explore the effectiveness of large kernel convolution in lightweight CNN-based SR models and achieve impressive SR performance, but there are still some limitations. First, the aggregation of non-local and local features plays a crucial role in model performance, but the fusion strategy still needs to be further studied. Second, due to time and computational constraints, this paper does not explore even larger convolution kernel sizes, such as 51 or beyond. Meanwhile, the reason why the SR performance declines when the kernel size exceeds 31 has not been fully analyzed. In future work, we will investigate these issues more thoroughly. Specifically, we will conduct a deeper analysis into why performance declines with larger kernel sizes and explore potential improvements. Moreover, we will continue to explore the application of large convolution kernels in lightweight SR models to further enhance their performance and efficiency.
\section{Conclusion}
\label{sec:conclusion}
In this paper, we propose LKFMixer, a pure CNN-based model that leverages large convolution kernels for efficient image SR. By utilizing coordinate decomposition and partial channel convolution, we successfully increase the kernel size to 31, significantly improving SR performance through enhanced non-local feature extraction, while maintaining lower model complexity and faster inference speeds. Extensive experiments demonstrate that the proposed method achieves SOTA SR performance and superior visual quality, with much faster inference speed compared with Transformer-based models. We hope our study will promote the research of large convolution kernels in lightweight SR models.

\bibliographystyle{IEEEtran}
\bibliography{References}

%

\end{document}